\documentclass[12pt]{article}
\usepackage{pdflscape}
\usepackage[T1]{fontenc}
\usepackage{geometry}
\usepackage{graphicx}
\usepackage{mathtools}
\usepackage{amsmath}
\usepackage{amssymb} 
\usepackage{dsfont} 
\usepackage[small]{caption}
\usepackage{mathrsfs}
\usepackage{upgreek}
\usepackage{eqparbox}
\usepackage{braket}
\usepackage{xfrac}
\usepackage{xspace}
\usepackage{booktabs}
\usepackage{subcaption}
\usepackage{letltxmacro}
\usepackage{multirow}
\usepackage{xcolor}
\usepackage{pgfornament}
\usepackage{tikz}
\usepackage{tikz-feynman}
\usepackage{slashed} 
\tikzfeynmanset{compat=1.1.0}
\usepackage{cancel}
\usepackage{float} 
\restylefloat{table}
\usepackage{dashrule}
\usepackage{longtable}
\usepackage{array}
\usepackage[numbers, compress, elide]{natbib}
\usepackage[no-natbib-sort]{jheppub}
\usepackage{comment}
\usepackage{soul}
\usepackage{cleveref} 
\makeatletter
\g@addto@macro\bfseries{\boldmath}
\makeatother
\allowdisplaybreaks
\newcommand{\Vev}{{v}} 
\newcommand{\M}{\mathcal{M}}
\newcommand{\A}{\mathcal{A}}

\newcommand{\cO}{\mathcal{O}}

\newcommand{\spu}{\lambda_H}
\newcommand{\sput}{{\widetilde{\lambda}_H}}

\newcommand{\Gew}{\rm{G}_{{\mathrm{EW}}}}

\newcommand{\beq}{\begin{equation}}
\newcommand{\eeq}{\end{equation}}

\newcommand{\anb}[1]{\langle#1\rangle}
\newcommand{\sqb}[1]{[#1]}
\newcommand{\asb}[1]{\langle#1]}
\newcommand{\sab}[1]{[#1\rangle}

\newcommand{\nn}{\nonumber}

\title{The Spurion Massive EFT}

\author[a]{Julian~L.~Northey,}
\author[a]{Yael~Shadmi,}
\author[a,b]{Yotam~Soreq,}
\author[a]{Daiki~Ueda}
\affiliation[a]{Physics Department, Technion - Israel Institute of Technology, Haifa 3200003, Israel}
\affiliation[b]{Theoretical Physics Department, CERN, 1211 Geneva 23, Switzerland}
\emailAdd{julian@campus.technion.ac.il}
\emailAdd{yshadmi@physics.technion.ac.il}
\emailAdd{soreqy@physics.technion.ac.il}
\emailAdd{daiki.ueda@campus.technion.ac.il}

\preprint{CERN-TH-2026-054} 

\abstract{ 
We use the amplitude formulation of the SMEFT to introduce a spurion analysis of the SMEFT low-energy amplitudes in terms of the Higgs VEV.
Each SMEFT contact-term is given as a sum of a few  spurion structures, whose number depends on the electroweak charges of the external legs.
The coefficients of these structures involve singlet combinations of Higgses from higher-order SMEFT contributions.
We use this to derive the spurion expansions of the $W$- and $Z$-boson masses and mixing, and their three-point couplings to fermions.
The textures of these couplings are saturated by the dimension-eight SMEFT.
Our analysis can be generalized to higher-point amplitudes and nonzero Yukawa couplings.
}

\begin{document}
\titlepage
\maketitle
\flushbottom

\section{Introduction}
\label{sec:1-Intro}

What is the SMEFT? 
From a bottom-up perspective, the Wilson coefficients of the Standard Model Effective Field Theory~(SMEFT)~\cite{Buchmuller:1985jz,Jenkins:2009dy,Grzadkowski:2010es} are unknown, 
so the SMEFT is determined by its global symmetry, and in particular, the $\Gew=$ SU(2)$_W\times $U(1)$_Y$ symmetry broken by a doublet-Higgs vacuum expectation value~(VEV). 
The low-energy observables predicted by the SMEFT should therefore be amenable to a spurion analysis with the small parameter being the Higgs VEV over the SMEFT scale, $\langle H\rangle/\Lambda\propto v/\Lambda$.
The only ``little hitch'' of course is that the broken symmetry is actually a {\it local} symmetry.
An amplitude-based construction of the SMEFT~\cite{Shadmi:2018xan,Ma:2019gtx,Aoude:2019tzn,Durieux:2019eor,Durieux:2019siw,Durieux:2020gip,Jiang:2020mhe,Dong:2021vxo,AccettulliHuber:2021uoa,Balkin:2021dko,Liu:2023jbq,Goldberg:2024eot,Dong:2025wvf} circumvents this difficulty, since the local symmetry is emergent. The amplitudes are completely dictated by the global  internal $\Gew$  symmetry in addition to spacetime symmetry.

In this paper, we introduce an amplitude-based Higgs-spurion analysis for the SMEFT, with the low-energy SMEFT amplitudes written as expansions in the Higgs VEV.
For each low-energy amplitude, the expansion involves a finite number of distinct spurion structures, corresponding to the $\Gew$-irreducible representations dictated by the relevant external legs. 
As a first application and demonstration of the method, we consider the most minimal, and yet most significant, electroweak precision observables, namely the three-point gauge-fermion amplitudes which determine $Z$- and $W$-pole observables 
(see, e.g., Refs.~\cite{Falkowski:2014tna, Efrati:2015eaa,Dawson:2019clf,Peskin:2020yqm}).
These will be particularly relevant for the Tera-$Z$ program at the FCCee \cite{Blondel:2019jmp,Agapov:2022bhm,Maura:2024zxz,Allwicher:2024sso}. 
We neglect fermion Yukawas, so we will only be concerned with the effects of vector masses, but our analysis straightforwardly generalizes to fermion masses, as well as to higher-point amplitudes.
We will work directly with an amplitude formulation of the SMEFT 
\cite{Shadmi:2018xan,Ma:2019gtx,Aoude:2019tzn,Durieux:2019eor,Durieux:2019siw,Durieux:2020gip,Jiang:2020mhe,Dong:2021vxo,AccettulliHuber:2021uoa,Balkin:2021dko,Liu:2023jbq,Goldberg:2024eot,Dong:2025wvf}, for which the spurion analysis is simple. 

The fact that the symmetry is local manifests itself in the gauge coupling.
Understanding how to handle the gauge coupling is the crucial ingredient in organizing the Higgs-spurion expansion.
In the Lagrangian formulation of the SMEFT, gauging the symmetry translates to replacing derivatives by covariant derivatives.
In the amplitude formulation, it is associated with the massless three-point gauge amplitudes.
In particular, the Higgs-vector gauge amplitude, $\A\left(H^\dagger HV \right)$, which we will refer to as $HHV$ for brevity, is the only  ingredient required for our analysis in addition to the symmetry breaking pattern.
There are two relevant consequences of the $HHV$ gauge coupling.
First, some gauge bosons get mass via mixing with the Goldstones. 
Second, as a result, different high-energy amplitudes unify into a single low-energy amplitude.
Indeed, from an amplitude perspective, the Higgs mechanism can be viewed as an IR unification of UV amplitudes~\cite{Arkani-Hamed:2017jhn}.
This was made concrete in Ref.~\cite{Balkin:2021dko}:
Each low-energy amplitude is generated by an infinite number of high-energy amplitudes with additional soft Higgs legs. 
The low-energy Higgsed amplitudes can then be obtained by matching to their high-energy origins~\cite{Balkin:2021dko,Liu:2023jbq,Goldberg:2024eot} (see also Refs.~\cite{Bachu:2019ehv,Bachu:2023fjn}).

In particular, the low-energy $Z$-fermion, or $W$-fermion, amplitudes have two origins: 
the high-energy gauge-fermion amplitudes, which contribute to the transverse polarization, and the fermion-Higgs amplitudes, which generate the longitudinal polarization. 
A single insertion of the  $HHV$ coupling fuses these together into a single amplitude.
Additional insertions of the $HHV$ coupling at tree-level merely modify the propagators and external lines from massless to massive.
It is therefore convenient to formulate the spurion expansion for small $g$, such that the masses are parametrically smaller than the VEV.
The two relevant scales in the problem, in addition to the EFT cutoff $\Lambda$, are then the Higgs VEV $v\ll\Lambda$, and the vector masses $m\sim g v< v$, where $g$ is the relevant gauge coupling. 
At scales below $v$ and above $m$, the SMEFT amplitudes are massless amplitudes and their structure can be determined by a standard spurion analysis in the Higgs VEV. 
At the scale $m$, these high-energy amplitudes can be matched to the massive amplitudes.
In practice, this scale separation is not crucial, since our focus is the SMEFT expansion, for which we only need the leading order in the gauge coupling. 
In this paper, we are only interested in a spurion analysis which gives the SU(2)$_W$ texture of the low-energy three-point amplitudes. 
In Ref.~\cite{OSH}, we use on-shell Higgsing to relate the low-energy amplitudes to the SMEFT contact-term amplitudes.

Our method can be readily generalized to four- and higher-point SMEFT amplitudes.
The SMEFT predictions for low-energy observables are given as a double expansion in derivatives, or equivalently, momenta, and in the Higgs VEV. 
The low-energy amplitudes we study here only feature the latter expansion, because the three-point kinematics is trivial. 
Higher-point amplitudes are given as a series expansion in kinematic structures. The coefficients of these different kinematic structures will involve different spurion expansions.
All-orders results in the Higgs VEV expansion for the fermion-vector amplitudes we consider here were derived in Ref.~\cite{Helset:2020yio} based on an analysis of the field-space geometry of the SMEFT Lagrangian~\cite{Alonso:2015fsp, Alonso:2016btr,Alonso:2016oah} (see also \cite{Corbett:2021cil, Martin:2023fad}).
Our approach is in some sense complementary, since it mainly relies on the $\Gew$ internal symmetry.
The SMEFT contact-term amplitudes are products of kinematic structures times $\Gew$ structures, which are overall (anti)symmetric under the exchange of identical (fermion) boson legs. For each kinematic structure, a standard group theory analysis determines the $\Gew$ structure of the amplitude.
The results we find are consistent with those of Ref.~\cite{Helset:2020yio}.

This paper is organized as follows. 
In Section~\ref{sec:three_pt}, we present the low-energy three-point amplitudes of interest.
The spurion expansions of the relevant SMEFT four-point amplitudes are discussed in Section~\ref{sec:four_point}, while the spurion expansion of the Higgs-gauge coupling is presented in Section~\ref{sec:massgen1}, where we also derive the resulting electroweak-vector spectrum. 
We then combine these results to present the low-energy couplings and discuss their implications in Section~\ref{sec:LE_coup}.
We conclude in Section~\ref{sec:conclusions}.
Further details are supplied in the Appendices.

\section{Low-energy amplitudes}
\label{sec:three_pt}

We use $V^I$ to denote the physical massive spin-1 bosons, with $V^{I=\pm}=W^\pm$, $V^{I=0}=Z$.
$F=Q, L$ denote fermion SU(2)$_W$-doublet quarks and leptons, and $f=u, d, e$ denote SU(2)$_W$ singlet fermions. 
We use $i,j=1,2$ to denote SU(2)$_W$ (anti-)fundamental indices, and $a=1,2,3$ for SU(2)$_W$ adjoint indices.
Note that, since we ignore the fermion Yukawa couplings, the fermions remain massless and there is no change of basis involved in their definition.

The low-energy amplitudes of interest feature a single kinematic structure each,
\begin{align}
    \begin{aligned}
    \label{eq:lowE}
    \M\left({\bar{F}}^i(p_1), {F^\prime}_j(p_2), V^I(p_3)\right)
    &= 
    (C^{I}_{F F^\prime})_i\,^j \, M_L(1,2,3)\,,
    \\
    \M\left(\bar{f}(p_1),f^\prime(p_2), V^I(p_3)\right)
    &=
    C^{I}_{ff^\prime}\, M_R(1,2,3)\,,
    \end{aligned}
\end{align}
where $M_L(1,2,3)$ and $M_R(1,2,3)$ contain the kinematic structures,
\begin{align}
   \begin{aligned} 
   M_L(1,2,3) 
   &=
   \bar v(p_1)\slashed {\epsilon}(p_3)P_L u(p_2)
   = 
   -\sqrt{2}\,\frac{[1{\bf3}]\anb{2{\bf3}}}{M_V}\,,
   \\
   M_R(1,2,3) 
   &=
   \bar v(p_1)\slashed{\epsilon}(p_3)P_R u(p_2)
   =
   -\sqrt{2}\,\frac{\anb{1{\bf3}}[2{\bf3}]}{M_V} \, ,
   \end{aligned}
\end{align}
with the Dirac spinors $u$ and $v$ denoting the particle and antiparticle solutions of the Dirac equation, respectively, $\epsilon$ the polarization vector, and $M_V$ the mass of $V^I$
(see Appendix~\ref{app:kin} for further details).
We emphasize that these are the physical masses of the on-shell vector bosons. 
Here and throughout, we take all momenta to be incoming.

In addition, the photon amplitudes are,
\begin{align}
    \begin{aligned}
    \label{eq:photon_coup}
    \A\left(\bar{F}^i(p_1), F_j(p_2), \gamma_+(p_3)\right)
    &=
    -\sqrt{2}\,  eQ_{F_i}\,\delta_i\,^j\, \frac{[13]^2}{[12]}\, ,
    \\
    \A\left(\bar{f}(p_1),f(p_2), \gamma_+(p_3)\right)
    &=
    \sqrt{2}\,eQ_f\, \frac{[23]^2}{[12]},
    \end{aligned}
\end{align}
where we wrote the spinor structures for positive helicity photons ($+$).
Recall that on-shell massless amplitudes with three external legs can be written for complex momenta.
Here and throughout, we use $\M$ for massive amplitudes and $\A$ for massless amplitudes. 
Our conventions mostly follow those of Ref.~\cite{Balkin:2021dko}.
We neglect Yukawa couplings and chirality-violating dipole amplitudes.

The transverse components of low-energy massive SMEFT amplitudes arise from massless vector amplitudes while their longitudinal components are generated by scalar amplitudes.
The low-energy amplitudes can be obtained by matching either the longitudinal or the transverse components~\cite{Balkin:2021dko}. 
(See Appendix~\ref{app:kin} for a brief review of the matching.)
The required SMEFT inputs for determining the low-energy amplitudes of Eq.~\eqref{eq:lowE} are therefore
(1)~the SMEFT fermion-fermion-Higgs-Higgs amplitudes, and
(2)~the SMEFT-corrected $HHV$ coupling,
which determines the vector masses.

\section{Deriving the coefficients of the low-energy amplitudes as spurion expansions} 
\label{sec:four_point}

Consider first the fermion-fermion-scalar-scalar amplitude contributing to the low-energy amplitudes.
For zero gauge couplings and nonzero Higgs VEV, the relevant  SMEFT amplitude  is,
\begin{align}
    \label{eq:ffGh}
     \A \left( \bar{\psi}(p_1), \psi^\prime(p_2) , G(p_3),h(p_4)\right) \, ,
\end{align}
where $G$ stands for any of the Goldstones ($G^0,\, G^\pm$) and $h$ is the radial Higgs mode.
This broken-phase amplitude can be covariantized with respect to the full $\Gew$ symmetry group using the Higgs VEV as a spurion, and rewriting the Goldstone and radial modes in terms of Higgs doublets,
\begin{align}
    h=(H^{\dagger\,2}+ H_2)/\sqrt2 \, ,
    \quad
    G^0=i(H^{\dagger\,2}- H_2)/\sqrt2\, , 
    \quad
    G^+=H_1\, ,
    \quad
    G^-=H^{\dagger\,1} \, .
\end{align}
For left-handed fermions, one relevant amplitude is then,
\begin{align}
    \label{eq:ffHHgeneral}
    \A \left( \bar{\psi}(p_1), \psi^\prime(p_2) , H^\dagger(p_3),H(p_4) \right)
    =
    \frac{\hat{c}_{\psi \psi^\prime HH}}{2}\, \frac{\sab{1(3-4)2}}{\Lambda^{2}} 
    =  
    \hat{c}_{\psi \psi^\prime HH}\, \frac{\sab{132}}{\Lambda^{2}}\,,
\end{align}
where to get the final expression we used momentum conservation.
For right-handed fermions $\sab{132}$ is replaced by $\langle132]$.
Here and throughout, hats denote the renormalized couplings, obtained by removing the appropriate factors of wave-function renormalizations~(WFRs).
To combine the different amplitudes correctly into massive low-energy amplitudes, we need to keep track of the WFRs.
We therefore define,
\begin{align}
    \label{eq:chatdefine}
   (c_{\psi \psi^\prime HH})_{i\ k}^{\ j \ \ell} 
   &\equiv 
   (\hat{c}_{\psi \psi^\prime HH})_{i\ m}^{\ j \ n}\,  (X^\dagger)_k\,^m\, X_n\,^\ell \,,
\end{align}
where the  matrix $X$ is the square root of the scalar WFR, and $i,j,k,\ell$ are the doublet indices of $\bar{\psi}, \psi^\prime , H^\dagger,H$ respectively.
In principle, we could similarly isolate the fermion WFRs, but these will play no role in the following.
Note that $(c_{\psi \psi^\prime HH})_{i\ k}^{\ j \ \ell}$ includes the full WFRs on the 
scalar legs and has well-defined properties under $\Gew$. 

Since we are considering the broken-phase theory, amplitudes featuring two $H$'s or two $H^\dagger$'s are also allowed, in addition to the amplitude of Eq.~\eqref{eq:ffHHgeneral}.

\subsection{Spurion structure of SMEFT amplitudes}
\label{sec:SMEFT-spurion}

We can now determine the coefficients $c_{\psi \psi^\prime HH}$ as expansions in the Higgs VEV, under the assumption of zero gauge couplings.
Since each additional Higgs leg comes with a factor of $1/\Lambda$, the spurion is $\langle H\rangle/\Lambda$ with $\langle H_2\rangle=v/\sqrt{2}$.
Because of the global symmetry, only even numbers of Higgses can appear, so it will be convenient to define the spurion,
\begin{align}
    \label{eq:spu}
    \spu^a 
    \equiv 
    2\,\langle H^\dagger \sigma^aH\rangle/\Lambda^2 \,,
\end{align}
which is an SU(2)$_W$ triplet of zero hypercharge.
Here and throughout, $\sigma^a$ denotes the Pauli matrices.
Substituting the Higgs VEV,
\begin{align}
    \spu^a = -\delta^{a3} \, v^2/\Lambda^2 \,.
\end{align}
In principle, the spurion $\sput^a\equiv 2\,\langle H^\dagger \sigma^a \widetilde H\rangle/\Lambda^2$, with $\widetilde{H}_i = (i\sigma^2)_{ij} H^{\ast\,j}$, which is a triplet of hypercharge $-1$, may also appear. 
As we will see, however, it is not required for the amplitudes of interest here.
No other combinations of Higgses are relevant.
Antisymmetric combinations of two $H$ fields (or two $\widetilde H$ fields) vanish;  $\langle H^\dagger \widetilde{H}\rangle$, $\langle \tilde{H}^\dagger H\rangle$,  and $\spu^a \sput^a$ are all zero. 
Thus, there are no $\mathrm{SU}(2)_W$-singlet spurions with nonzero hypercharge.
Obviously, the singlet combination $\langle H^\dagger H\rangle$ does appear: 
all the coefficients below are infinite series expansions in 
$2\,\langle H^\dagger H\rangle/\Lambda^2=v^2/\Lambda^2$.~\footnote{One can similarly treat Higgs insertions in the SM by defining the dimensionless spurion $\langle H\rangle/v$.}

Consider first the left-handed fermion amplitude, $\A \left( \bar{F}^{i}(p_1), F'_{j}(p_2) , H^{\dagger\,k}(p_3),H_{\ell}(p_4) \right)$ of Eq.~\eqref{eq:ffHHgeneral}, where $F$ and $F'$ may denote different generations.
Generically, for four arbitrary SU(2)$_W$-doublets, the amplitude can feature six independent coefficients corresponding to the six associated independent irreps of $\Gew$,
\begin{align} 
    \label{eq:ffssLH1group-try}
   (c_{FF'HH})_{i\ k}^{\ j \ \ell} 
    &=
     c_{FF'1} \,\delta_{i}^{\,\,\,j} \delta_{k}^{\,\,\,\ell} 
    +c_{FF'2}\,(\sigma^a)_{i}^{\,\,\,j} (\sigma^a)_{k}^{\,\,\,\ell}
    + c_{FF'3}^a \,(\sigma^a)_{i}^{\,\,\,j} \delta_{k}^{\,\,\,\ell} 
    \nn \\
    &\quad+  c_{FF'4}^c \, \varepsilon^{abc}
    (\sigma^a)_{i}^{\,\,\,j} (\sigma^b)_{k}^{\,\,\,\ell} +c_{FF'5}^b \,\delta_{i}^{\,\,\,j} (\sigma^b)_{k}^{\,\,\,\ell} 
    + 
    c_{FF'6}^{ab}
    \,(\sigma^a)_{i}^{\,\,\,j} (\sigma^b)_{k}^{\,\,\,\ell} 
    \,.
\end{align}   
The coefficients $c_{FF'1}$ through $c_{FF'6}$ are $\mathrm{SU}(2)_W$ tensor couplings, with repeated adjoint indices summed over. 
$c_{FF'6}^{ab}$ is symmetric.
In the SMEFT, these couplings arise from higher-dimensional operators and are expressed in terms of the single spurion $\spu^a$,
\begin{align}           
    &c_{FF'1}=c_{FF'1}^{(6)} \,,\quad 
   c_{FF'2}=c_{FF'2}^{(6)} \,,
   \notag \\
   &c^a_{FF'3}=c_{FF'3}^{(8)}\,\spu^a \,, \quad
    c^a_{FF'4}=c_{FF'4}^{(8)}\,\spu^a \,,\quad 
    c^a_{FF'5}=c_{FF'5}^{(8)}\,\spu^a \,,\quad 
    \notag\\
    &c^{ab}_{FF'6}=c_{FF'6}^{(10)}\,\spu^a\spu^b\,, 
\end{align}   
where superscripts indicate the dimension at which each term can appear.

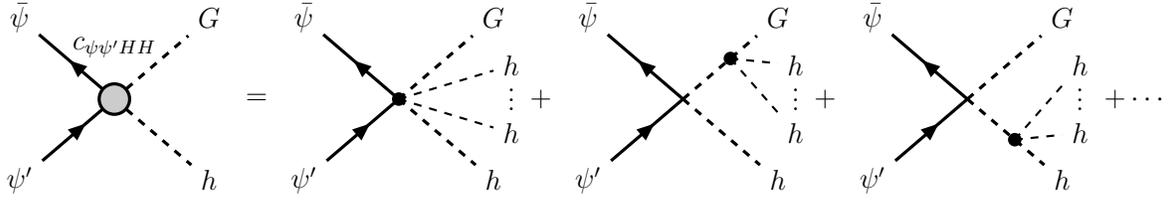
\begin{figure}[t]
\centering
\begin{tikzpicture}[scale=0.9, transform shape]
\begin{feynman}
  \vertex (a1) at (1.0,-1.2) {$\psi'$};
   \vertex (b1) at (1.0,1.2) {$\bar{\psi}$};
    \vertex (e1) at (3.8,1.2) {$G$};
    \vertex (g1) at (2.4,0) ;
    \vertex (h1) at (3.8,-1.2) {$h$} ;
    \vertex (i1) at (3.8,0.5) ;
    \vertex (j1) at (3.8,-0.5) ;
    \vertex (k1) at (3.0,0) ;
    \vertex (l1) at (3.1,0.6) ;
    \vertex (m1) at (3.1,-0.6) ;
     \vertex [right=4.2cm of a1] (a2) {$\psi'$};
     \vertex [right=4.2cm of b1] (b2) {$\bar{\psi}$};
    \vertex [right=4.2cm of e1] (e2) {$G$};
    \vertex [right=4.2cm of g1] (g2) ;
    \vertex [right=4.2cm of h1] (h2){$h$} ;
    \vertex [right=4.2cm of i1] (i2){$h$} ;
    \vertex [right=4.2cm of j1] (j2){$h$} ;
    \vertex [right=4.2cm of k1] (k2) ;
    \vertex [right=4.2cm of l1] (l2) ;
    \vertex [right=4.2cm of m1] (m2) ;
    \vertex [right=4.2cm of a2] (a3) {$\psi'$};
     \vertex [right=4.2cm of b2] (b3) {$\bar{\psi}$};
    \vertex [right=4.2cm of e2] (e3) {$G$};
    \vertex [right=4.2cm of g2] (g3) ;
    \vertex [right=4.2cm of h2] (h3){$h$} ;
    \vertex [right=4.2cm of i2] (i3){$h$} ;
    \vertex [right=4.2cm of j2] (j3){$h$} ;
    \vertex [right=4.2cm of k2] (k3) ;
    \vertex [right=4.2cm of l2] (l3) ;
    \vertex [right=4.2cm of m2] (m3) ;
    \vertex [right=4.2cm of a3] (a4) {$\psi'$};
     \vertex [right=4.2cm of b3] (b4) {$\bar{\psi}$};
    \vertex [right=4.2cm of e3] (e4) {$G$};
    \vertex [right=4.2cm of g3] (g4) ;
    \vertex [right=4.2cm of h3] (h4){$h$} ;
    \vertex [right=4.2cm of i3] (i4){$h$} ;
    \vertex [right=4.2cm of j3] (j4){$h$} ;
    \vertex [right=4.2cm of k3] (k4) ;
    \vertex [right=4.2cm of l3] (l4) ;
    \vertex [right=4.2cm of m3] (m4) ;
    
\node[circle, fill=black, inner sep=0pt, minimum size=2.0mm] (bk) at (g2) {};
\node[circle, fill=black, inner sep=0pt, minimum size=2.0mm] (bk) at (l3) {};
\node[circle, fill=black, inner sep=0pt, minimum size=2.0mm] (bk) at (m4) {};

  \diagram*{
    (a1) -- [fermion, line width=1.2pt] (g1) -- [fermion, line width=1.2pt] (b1),
      (g1) -- [scalar, line width=1.2pt] (e1),
      (g1) -- [scalar, line width=1.2pt] (h1),
    (a2) -- [fermion, line width=1.2pt] (g2) -- [fermion, line width=1.2pt] (b2),
      (g2) -- [scalar, line width=1.2pt] (e2),
      (g2) -- [scalar, line width=1.2pt] (h2),
      (g2) -- [scalar, line width=0.8pt] (i2),
      (g2) -- [scalar, line width=0.8pt] (j2),
      (a3) -- [fermion, line width=1.2pt] (g3) -- [fermion, line width=1.2pt] (b3),
      (g3) -- [scalar, line width=1.2pt] (e3),
      (g3) -- [scalar, line width=1.2pt] (h3),
      (l3) -- [scalar, line width=0.8pt] (i3),
      (l3) -- [scalar, line width=0.8pt] (j3),
      (a4) -- [fermion, line width=1.2pt] (g4) -- [fermion, line width=1.2pt] (b4),
      (g4) -- [scalar, line width=1.2pt] (e4),
      (g4) -- [scalar, line width=1.2pt] (h4),
      (m4) -- [scalar, line width=0.8pt] (i4),
      (m4) -- [scalar, line width=0.8pt] (j4)
  };

  \node at ($(h1)!0.5!(b2) + (0,0)$) {$=$};
  \node at ($(h2)!0.5!(b3) + (0,0)$) {$+$};
  \node at ($(h3)!0.5!(b4) + (0,0)$) {$+$};
  \node at ($(h3)!0.5!(b4) + (4.6,0)$) {$+\cdots$};
  \node at ($(i2)!0.5!(j2)+ (0, 1.0 mm)$) {$\vdots$};
  \node at ($(i3)!0.5!(j3)+ (0, 1.0 mm)$) {$\vdots$};
  \node at ($(i4)!0.5!(j4)+ (0, 1.0 mm)$) {$\vdots$};
  \node[circle,
  draw=black,
  fill=gray!40,
  fill opacity=1,
  draw opacity=1,
  line width=1.2pt,
  inner sep=0pt,
  minimum size=4.5mm,
  preaction={fill=white}] at (g1) {};
  \node[above=4.7mm of g1] {$c_{\psi\psi'HH}$};
\end{feynman}
\end{tikzpicture}
\vspace{-14pt}
\caption{Origin of $c_{\psi \psi'HH}$ from multiple soft Higgs insertions (see text for details).
}
\label{fig:four_pt}
\end{figure}

The origin of this spurion expansion is clear.
The massless amplitudes of the unbroken theory respect the full global symmetry, with $\Gew$ broken by the Higgs VEV.
The broken-phase amplitude Eq.~\eqref{eq:ffHHgeneral} is obtained from the soft-Higgs limits of an infinite set of unbroken amplitudes, with each soft Higgs leg suppressed by $v/\Lambda$.
The $\Gew$-indices of the amplitude Eq.~\eqref{eq:ffHHgeneral} may be contracted with the indices of these soft 
Higgses\footnote{For hypercharge this means that the overall hypercharge of the low-energy amplitude, if nonzero, is compensated by the overall hypercharge of the soft Higgses.}.
Summing the contributions of all amplitudes with extra Higgs legs generates 
$(c_{FF'HH})_{i\ k}^{\ j \ \ell}$ (see Fig.~\ref{fig:four_pt}), 
which has well-defined properties under the unbroken $\Gew$ symmetry and hence a well-defined spurion expansion.
As mentioned above, $(c_{FF'HH})_{i\ k}^{\ j \ \ell}$ includes contributions from all Higgs insertions on the external legs, and these 
generate the full WFR, namely $X^\dagger X$.
Note that the SMEFT-generated WFRs are only multiplicative factors. 
Nothing gets mass at this point, so schematically, SMEFT contact-terms give $p^2$ corrections on external legs and these cancel against $1/p^2$ propagators.

For $F=F'$, the couplings $c_{FF'1}^{(6)},\ldots, c_{FF'6}^{(10)} $ are all real.
To see this consider the complex-conjugated amplitude, which can be related to the original amplitude by exchanging the appropriate external legs (for details see, e.g., Ref.~\cite{Goldberg:2024eot}).
As mentioned above, each of these couplings is an infinite sum over SMEFT Wilson coefficients involving  singlet combinations of Higgs spurions
$2\, \langle H^\dagger H\rangle/\Lambda^2= \Vev^2/\Lambda^2$. 

In fact, only four of the structures above lead to independent contributions to the fermion-vector three-point couplings.
The fermion-vector amplitudes are obtained by taking the soft limit of the Higgs field, $H^{\dagger\,2}$ or $H_2$, whose direction in $\Gew$-space are fully correlated  with the spurion direction (i.e., the Higgs VEV).
Specifically, for $G^0$, only $k=\ell=2$ contributes, whereas for $G^{\pm}$, only $k\neq \ell$ contributes to the three-point amplitudes.
When focusing on these three-point amplitudes, the relation $\lambda_H^a \propto \delta^{a3}$ implies that the couplings $c^{(8)}_{FF'5}$ and $c^{(10)}_{FF'6}$ are redundant. 
They can be traded for $c_{FF'1}^{(6)}$ and for $c_{FF'3}^{(8)}$, respectively.
For example, 
$c^{(8)}_{FF' 5}\,\delta_i\,^j (\sigma^b)_{k}\,^\ell \lambda^b_H=(v^2/\Lambda^2)\, c^{(8)}_{FF' 5}\, \delta_i\,^j \delta_{k}\,^\ell\propto c_{FF'1}^{(6)}\,\delta_i\,^j \delta_{k}\,^\ell$ for $k=2$ or $\ell=2$.
More generally, all six couplings can contribute. 
Thus for instance, they contribute independently to $\bar{F}F W^+W^-$ and $\bar{F}F ZZ$ (which require an additional factor of $(p_3+p_4)^2$ in the kinematic structure). 
They would also give independent contributions in the presence of additional sources of $\Gew$ breaking, such as a Higgs triplet.

To summarize, the required input for the fermion-vector coupling in the SMEFT is then,
\begin{align} 
    \label{eq:ffssLH1group}
   (c_{FF'HH})_{i\ k}^{\ j \ \ell}
    =\, & 
    c_{FF'1}^{(6)} \,\delta_{i}^{\,\,\,j} \delta_{k}^{\,\,\,\ell} 
    +c_{FF'2}^{(6)}\,(\sigma^a)_{i}^{\,\,\,j} (\sigma^b)_{k}^{\,\,\,\ell}\,\delta^{ab}\notag \\
    &+ c_{FF'3}^{(8)}\, \spu^a\,(\sigma^a)_{i}^{\,\,\,j} \delta_{k}^{\,\,\,\ell} 
    + c_{FF'4}^{(8)} \, \varepsilon^{abc} \spu^c
    (\sigma^a)_{i}^{\,\,\,j} (\sigma^b)_{k}^{\,\,\,\ell} 
    \,. 
\end{align}   
In Section~\ref{sec:LE_coup}, we will use Eq.~\eqref{eq:ffssLH1group} to determine $C^I_{FF^\prime}$ of Eq.~\eqref{eq:lowE}. 
Note that the amplitude $\A( \bar F^{i},F'_{j},H_k,H_{\ell})$ does not add new information, because it arises, at leading order, from the unbroken six-point amplitude $\A( \bar F^i,F'_j,H^{\dagger\,m}, H_k,H^{\dagger\,n}, H_\ell)$, which also generates the amplitude of Eq.~\eqref{eq:ffHHgeneral}, and $H$ and $H^\dagger$ yield the same spurion in the soft Higgs limit.
This can be checked explicitly: 
the single spurion structure allowed in 
$\A( \bar F^{i},F'_{j},H_k,H_{\ell})$ is $\varepsilon^{k\ell}\, (\sigma^a)_i\,^j \tilde{\lambda}^a_H$ with $\varepsilon_{k\ell}=-\varepsilon^{k\ell}$ and $\varepsilon_{21}=1$,
but this can be related to the $c_{FF'4}^{(8)}$ term of Eq.~\eqref{eq:ffssLH1group}, 
using $\varepsilon_{ij}=-i(\sigma^2)_{ij}$ and $\sigma^a \sigma^2=\delta^{a2}1+i\varepsilon^{a2b}\sigma^b$.

We can similarly obtain the spurion expansions for the remaining amplitudes.
For SU(2)$_W$-singlet fermions, 
\begin{align}
    \label{eq:ffssRH1-full}
     \A\left( \bar{f}(p_1), f'(p_2) , H^{\dagger\,k}(p_3),H_{\ell}(p_4)\right)
     =
    (\hat{c}_{ff'HH})_k\,^\ell
   \, \frac{\langle132]}{\Lambda^{2}}
    \,,
\end{align}
with $(c_{ff'HH})_{k}\,^\ell = (\hat{c}_{ff'HH})_{ m}\,^{n} (X^\dagger)_k\,^m X_n\,^\ell= c_{ff'1}^{(6)} \,\delta_{k}\,^\ell + c_{ff'2}^{(8)}\, \spu^a (\sigma^a)_k\,^\ell$. 
As in the left-handed fermion case, the general structure can be reduced to a single structure, and the relevant input for the right-handed fermion-vector coupling is
\begin{align}
    ({c}_{ff'HH})_{k}\,^\ell
    =
    c_{ff'1}^{(6)} \,\delta_{k}\,^\ell \,.\label{eq:ffssRH1}
\end{align}

For the right-handed quarks there is an additional amplitude coming from the following U(1)$_Y$ neutral combination
\begin{align}
    \label{eq:udssRH1-full}
    \A \Big( \bar{u}(p_1), d(p_2) ,H_k(p_3), H_{\ell}(p_4)\Big)
    = \frac12 \hat{c}_{ud}^{(6)} \,\varepsilon^{k\ell}\, \frac{\asb{1(3-4)2}}{\Lambda^{2}}=
    \hat{c}_{ud}^{(6)} \,\varepsilon^{k\ell}\, \frac{\asb{132}}{\Lambda^{2}} \, .    
\end{align}
In addition, the amplitude $\A ( \bar{u}, d , H^{\dagger\,k}, H_{\ell})$ yields non-vanishing structures of the form $(\sput^a)^\ast (\sigma^a)_k\,^\ell$.
However, these contributions can be absorbed into the coefficient $c^{(6)}_{ud}\varepsilon^{k\ell}=\hat{c}^{(6)}_{ud}\varepsilon^{mn} X_m\,^k X_n\,^\ell$. 
Note also that $c^{(6)}_{ud}$ is generally complex-valued.

The couplings above determine the couplings of the low-energy massive $\bar\psi\psi' V$ amplitudes.
They also determine some of the couplings appearing in the four-point amplitude $\bar\psi\psi' V h$.
These four-point amplitudes can feature several different kinematic structures, including the one we considered here, $[132\rangle$, which becomes ${\sqb{1 \mathbf{3}}\anb{2 \mathbf{3}}}$ in the low-energy $\bar{F} F' V h$ amplitudes.
The coefficients of $[13]\anb{23}$ are determined by the couplings above. 
For instance, in $\bar F F' V h$, the Wilson coefficient of $[13]\anb{23}$ is given by $c_{FFHH}$ for appropriate choices of indices. 
Thus, as expected, the SMEFT predicts relations between three-point and higher-point couplings.

\subsection{Restoring the gauge coupling}
\label{sec:gauge}
For zero gauge coupling and nonzero Higgs VEV, the theory is an interacting theory of massless
fermions, vectors, Goldstone bosons and the Higgs radial mode.
Turning the gauge coupling back on has several effects.
First, some of the amplitudes receive SM contributions at second order in the gauge couplings $g_1$ and $g_2$, arising from $B$- and $W$-exchange, respectively.
These are captured by the spurion expansions above. 
At tree-level, $W$-exchange generates $c_{FF'2}^{(6)}$ and $B$-exchange generates $c_{FF'1}^{(6)}$
(see Eq.~\eqref{eq:ffssLH1-sm} in the Appendix for the explicit expression).

Furthermore, there is now a $H H V$ three-point amplitude, which generates the vector mass and Goldstone-vector mixing. 
Matching the amplitudes of Section~\ref{sec:SMEFT-spurion} to the longitudinal components of the amplitudes~\eqref{eq:lowE} we can then derive the low-energy couplings $C^{I}_{FF^\prime}$, $C^{I}_{f f^\prime}$ of Eq.~\eqref{eq:lowE}.
Amplitudes with a $G^\pm$ Goldstone external leg determine the low-energy $W^\pm$ amplitudes, while amplitudes with a $G^0$ Goldstone external leg determine the low-energy $Z$ amplitudes.
So all we need to do is to choose the Higgs indices of the coefficients, e.g., $c_{FF'HH}$ and $c_{ff'HH}$ listed in Section~\ref{sec:SMEFT-spurion} to be $k=\ell=2$ to get the $Z$ couplings, and $k\neq \ell$ to get the $W$ couplings.
We do this in Section~\ref{sec:LE_coup}.
First, however, we determine the remaining low-energy parameters, namely the masses $M_W$ and $M_Z$, along with the mixing angle and $\rho$ parameter.

\section{Generating masses}
\label{sec:massgen1}

\subsection{Massive states from the $HHV$ coupling}
\label{sec:massive-states}

The source of vector masses is the $HHV$ three-point amplitude\footnote{Recall that the $H^\dagger HVV$ Feynman-rule vertex is not an independent input. 
It is determined from the three-point coupling and the gauge symmetry, and in particular can be rotated away, see, e.g., Ref.~\cite{Elvang:2015rqa}.}, which, for a positive-helicity vector is,
\begin{align}
    \label{eq:HHV}
    \A(H^{\dagger\, i}(p_1), H_j(p_2), V^{A}_+(p_3)) 
    =  
    (\hat{g}^A_H)_{i}\,^j (p_1-p_2)\cdot \epsilon_+(p_3)
    =
    -\sqrt{2}\,(\hat{g}^A_H)_{i}\,^j \, \frac{[13][23]}{[12]}\, ,
\end{align}
where $\epsilon$ denotes the polarization vector.
The index $A=a,4$ runs over the four massless gauge bosons, namely the fields with well-defined $\Gew$ transformation properties, $W^a, B$.
The amplitude in Eq.~\eqref{eq:HHV} is the SMEFT-corrected gauge coupling:
$\hat{g}^A_H$ is linear in the gauge coupling but includes all SMEFT corrections for nonzero $\Vev$.
We note that there is no SMEFT contact-term that generates a vector-scalar mixing for zero gauge coupling~\cite{OSH}.
Thus, the only source of the mass is Eq.~\eqref{eq:HHV}.
\begin{figure}[t]
\centering
   \begin{subfigure}[b]{0.3\textwidth}
    \centering
       \begin{tikzpicture}
  \begin{feynman}
    \vertex (b) at (1.0,0) {$V^A$};
    \vertex (e) at (2.4,1.5) {$h$};
    \vertex (g) at (2.4,0) ;
    \vertex (h) at (3.8,0) {$G$} ;

    \diagram* {
      (g) -- [photon, line width=1.2pt] (b),
      (g) -- [scalar, line width=1.2pt] (e),
      (g) -- [scalar, line width=1.2pt] (h)
    };
    \node[circle,
  draw=black,
  fill=gray!40,
  fill opacity=1,
  draw opacity=1,
  line width=1.2pt,
  inner sep=0pt,
  minimum size=4.5mm,
  preaction={fill=white}] at (g) {};
  \end{feynman}
\end{tikzpicture}
  \end{subfigure}
     \begin{subfigure}[b]{0.3\textwidth}
    \centering
       \begin{tikzpicture}
  \begin{feynman}
    \vertex (b) at (1.0,0) {$G$};
    \vertex (e) at (2.4,1.5) {$h$};
    \vertex (g) at (2.4,0) ;
    \vertex (h) at (3.8,0) {$V^B$};

    \diagram* {
      (g) -- [scalar, line width=1.2pt] (b),
      (g) -- [scalar, line width=1.2pt] (e),
      (g) -- [photon, line width=1.2pt] (h)
    };
    \node[circle,
  draw=black,
  fill=gray!40,
  fill opacity=1,
  draw opacity=1,
  line width=1.2pt,
  inner sep=0pt,
  minimum size=4.5mm,
  preaction={fill=white}] at (g) {};
  \end{feynman}
\end{tikzpicture}
  \end{subfigure}
  \begin{subfigure}[b]{0.3\textwidth}
    \centering
       \begin{tikzpicture}
  \begin{feynman}
    \vertex (b) at (1.0,0) {$V^A$};
    \vertex (e) at (2.4,1.5) {$h$};
    \vertex (f) at (3.6,1.5) {$h$};
    \vertex (g) at (2.4,0) ;
    \vertex (h) at (3.6,0) ;
    \vertex (i) at (5.0,0) {$V^B$};
   
    \diagram* {
      (g) -- [photon, line width=1.2pt] (b),
      (g) -- [scalar, line width=1.2pt] (e),
      (g) -- [scalar, edge label=$G$, line width=1.2pt] (h),
      (h) -- [scalar, line width=1.2pt] (f),
      (h) -- [photon, line width=1.2pt] (i)
    };
    
 \node[circle,
  draw=black,
  fill=gray!40,
  fill opacity=1,
  draw opacity=1,
  line width=1.2pt,
  inner sep=0pt,
  minimum size=4.5mm,
  preaction={fill=white}] at (g) {};
    \node[circle,
  draw=black,
  fill=gray!40,
  fill opacity=1,
  draw opacity=1,
  line width=1.2pt,
  inner sep=0pt,
  minimum size=4.5mm,
  preaction={fill=white}] at (h) {};

  \end{feynman}
\end{tikzpicture}
  \end{subfigure}
\caption{A single $HHV$ insertion with a soft Higgs $h$ converts an external vector $V$ into a Goldstone $G$ and vice versa; two such insertions generate an effective mass correction to the vector propagator.
The blob denotes the SMEFT-corrected gauge coupling (see Fig.~\ref{fig:diagram-eq2}).
}
\label{fig:mass}
\end{figure}
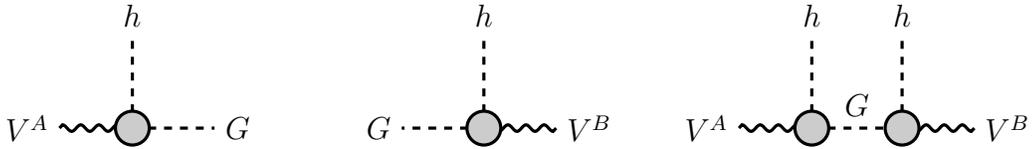

A single insertion of $HHV$, with one soft Higgs leg, flips an external vector into an external Goldstone and vice versa, see Fig.~\ref{fig:mass}, and unifies the scalar and vector amplitudes into a single massive amplitude. 
The derivation of the low-energy couplings therefore only relies on the inclusion of a single $HHV$ vertex.
Additional insertions of the $HHV$ coupling modify the propagators from massless to massive once the Higgs momenta are taken to be soft, see Fig.~\ref{fig:mass}.
Since the low-energy amplitudes of interest are three-point amplitudes, they do not feature any propagators, but we still need to consider multiple insertions of the $HHV$ coupling to determine the $W$ and $Z$ masses and mixing.
The mass-squared matrix is given by,
\begin{align}
    \label{eq:mass-sq}
    (M^2)^{AB} 
    = 
    \frac{v^2}{2} \left( (\hat{g}^{A}_H)^{\,\,\,i}_{2} (\hat{g}^{B}_H)^{\,\,\,2}_{i} + {\rm h.c.}
    \right) \,.
\end{align}
Thus, the $HHV$ coupling defines two vectors in the space $V^A=(W^a,B)$,
\begin{align}
    (\hat{g}^A_H)^{\,\,\,j=2}_{i=1}, 
    ~~~\text{and}~~~
    (\hat{g}^A_H)^{\,\,\,j=2}_{i=2}\,,
\end{align}
corresponding to the two massive eigenstates: 
the Goldstone corresponding to $i=1$ defines the $W$, while the Goldstone corresponding to $i=2$ defines the $Z$.

It is instructive to rewrite these combinations as normalized vectors, 
\begin{align}
    \tilde V_i^A
    \equiv  
    (\hat{g}^A_H)^{\,\,\,j=2}_{i}/N_i\,,
\end{align}
where the normalization $N_i$ is given by 
$N_i= \left[\sum_{A=1}^4 \left\vert(\hat{g}^A_H)^{\,\,\,j=2}_{i}\right\vert^2\right]^{1/2}$.
The mass-squared matrix can then be written as,
\begin{align}
  M^2
  = 
  \frac12\, v^2\, \sum_i N_i^2\, \left(\tilde V_i \tilde V^{\dagger\,i} +{\rm h.c.}\right)  \,.
\end{align}
Thus,  $\tilde V_i$ are eigenvectors of $M^2$ with eigenvalues $v^2(N_i)^2$, and the photon corresponds to the orthogonal combination (note that $M^2$ is rank 3).
These vectors give both the rotation to the mass basis and the masses, with $M_W=v\,N_1/\sqrt{2}$ and $M_Z=v\,N_2$.

\subsection{Spurion structure of the $HHV$ coupling}
\label{sec:SpuStrucuture}

\begin{figure}[t]
\centering
  \begin{tikzpicture}
\begin{feynman}
   \vertex (b1) at (1.0,0) {$V^A$};
    \vertex (e1) at (2.4,1.5) {$H_j$};
    \vertex (g1) at (2.4,0) ;
    \vertex (h1) at (3.8,0) {$H^{\dagger\,i}$} ;
    \vertex (i1) at (1.85,0) ;
    \vertex (j1) at (2.4,0.7) ;
    \vertex (k1) at (2.95,0) ;
     \vertex [right=4.7cm of b1] (b2) {$V^A$};
    \vertex [right=5.0cm of e1] (e2) {$H_j$};
    \vertex [right=5.0cm of g1] (g2) ;
    \vertex [right=5.3cm of h1] (h2){$H^{\dagger\,i}$} ;
    \vertex [right=4.9cm of i1] (i2) ;
    \vertex [right=5.0cm of j1] (j2) ;
    \vertex [right=5.1cm of k1] (k2) ;
    
    \node[circle, fill=black, inner sep=0pt, minimum size=2.0mm] (bi) at (i2) {};
\node[circle, fill=black, inner sep=0pt, minimum size=2.0mm] (bj) at (j2) {};
\node[circle, fill=black, inner sep=0pt, minimum size=2.0mm] (bk) at (k2) {};

\node[below=4.1mm of bi] {\footnotesize $K^{AB}$};
\node[right=4.9mm of bj] {\footnotesize $X_k\,^j$};
\node[below=4.1mm of bk] {\footnotesize $(X^\dagger)_i\,^\ell$};

  \diagram*{
    (g1) -- [photon, line width=1.2pt] (b1),
      (g1) -- [scalar, line width=1.2pt] (e1),
      (g1) -- [scalar, line width=1.2pt] (h1),
    (g2) -- [photon, line width=1.2pt] (b2),
      (g2) -- [scalar, line width=1.2pt] (e2),
      (g2) -- [scalar, line width=1.2pt] (h2),
  };

  \node at ($(h1)!0.5!(b2) + (0,0.6)$) {$=$};
  \node[circle,
  draw=black,
  fill=gray!40,
  fill opacity=1,
  draw opacity=1,
  line width=1.2pt,
  inner sep=0pt,
  minimum size=4.5mm,
  preaction={fill=white}] at (g1) {};
\end{feynman}
\end{tikzpicture}
\caption{A schematic picture of the $HHV$ three-point amplitude in Eq.~\eqref{eq:HHV}. }
\label{fig:diagram-eq2}
\end{figure}
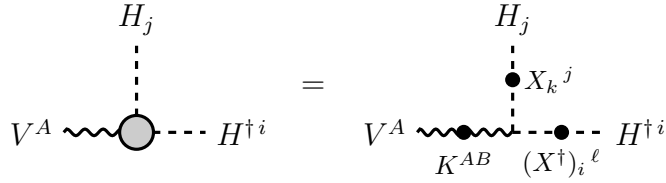
Next, we determine $(\hat{g}^A_H)^{\,\,\,j=2}_{i}$ in the SMEFT.
Since it is proportional to the gauge coupling, the SMEFT corrections to this coupling are multiplicative (see Fig.~\ref{fig:diagram-eq2}), 
\begin{align}
    \label{eq:renorm}
    (\hat{g}^A_H)_{i}\,^j
    = 
    X_{\ell}\,^{j} \,(X^\dagger)_{i}\,^{k} \,K^{AB}\,(g^B_H)_{k}\,^{\ell} \,,   
\end{align}
where
\begin{align}
\label{eq:gauge_coup}
    (g^b_H)_{k}\,^\ell 
    =   
    \frac{1}{2}\,g_2\, (\sigma^b)_{k}\,^\ell\, , 
    \qquad \qquad
    (g^4_H)_{k}\,^\ell 
    =   
    \frac{1}{2}\, g_1\,  \delta_{k}\,^\ell\,,
\end{align}
are the SM tree-level values.

The matrices $K$ and $X$ are the square-roots of the vector and scalar WFRs respectively.
We can write down the spurion expansion for the scalar and vector two-point functions, as we did for $c_{\psi\psi' HH}$.
It is more convenient however to do this directly for the square-roots of these matrices\footnote{Obviously, the square-root of the WFR is defined up to some matrix, which corresponds to a change of basis. 
In particular, by writing the spurion expansions for the square roots, we are implicitly taking this basis to be the unbroken basis, such that, e.g., $K$ has the indices $A,B$.},
\begin{align}
    \label{eq:K}
    K^{ab}
    &=
    \left(1+c_{WW1}^{(6)}\,\frac{v^2}{\Lambda^2}\right)\delta^{ab}+c_{WW2}^{(8)}\, \spu^a \spu^b\,,
    \nn\\
    K^{a4}
    &=
    K^{4a}
    = 
    c_{WB}^{(6)}\, \spu^a \, , 
    \\
    K^{44}
    &=
    1+c_{BB}^{(6)}\,\frac{v^2}{\Lambda^2} \, , 
    \nn
\end{align}
and 
\begin{align}
    \label{eq:X}
    X_i\,^j= 
    \left(1+c_{HH1}^{(6)}\,\frac{v^2}{\Lambda^2}\right)\, \delta_i\,^j 
    + c_{HH2}^{(6)}\, \spu^a (\sigma^a)_i\,^j\,,
\end{align}
where the superscripts indicate the operator dimension at which each coefficient can first appear in the SMEFT expansion.
Note that $K$ and $X$ are  Hermitian. Furthermore,  $X$ is diagonal.
This is a consequence of the unbroken U(1)$_{\rm{EM}}$ symmetry; 
in more general Higgsed theories, this matrix will be block diagonal, with the blocks corresponding to the unbroken subgroups.
The coefficients $c^{(6)}_{WW1}$, $c^{(8)}_{WW2}$, ..., and $c^{(6)}_{HH2}$, in Eqs.~\eqref{eq:K}-\eqref{eq:X} are the oblique parameters and receive contributions also from SM loops~\cite{Peskin:1991sw}.

In fact, the parametrization in Eq.~\eqref{eq:renorm}, together with Eqs.~\eqref{eq:gauge_coup}, \eqref{eq:K}, and~\eqref{eq:X}, is equivalent to the {\it most general} spurion expansion of $\hat{g}_H$,
\begin{align}
    \label{eq:Hgeneral}
    (\hat{g}^a_H)_{i}\,^j 
    &=
    c_{H W1}\, (\sigma^a)_{i}\,^j + 
    c_{HW 2}\, \spu^a \delta_{i}\,^j  +
    c_{H W 3}\,  \spu^a \spu^b (\sigma^b)_{i}\,^j\,, 
    \\
    (\hat{g}^4_H)_{i}\,^j 
    &=
    c_{H B1}\, \delta_{i}\,^j + c_{H B 2}\, \spu^a (\sigma^a)_{i}\,^j \,,
\end{align}
provided that the spurion coefficients $c_{H W1}$ to $c_{H B 2}$ vanish for zero gauge couplings.
Note that these coefficients are real-valued, because $\hat{g}^a_H$ is Hermitian (see Appendix~\ref{app:mass-spur-gen}).
The relations of these coefficients to the matrices $X$ and $K$ are given in Eq.~\eqref{eq:gen-xk-reln}. 

\subsection{Masses, mixings and custodial symmetry breaking}
\label{sec:mass-spur-XK}

As discussed in Sections~\ref{sec:massive-states} and~\ref{sec:SpuStrucuture}, the masses and mass states are defined by
\begin{align}
    (\hat{g}^A_H)_i\,^{j=2} 
    = 
   K^{AB}\, (X^\dagger \,g^B_H\,X)_{i}\,^2 \, .
\end{align}
Since $X$ is diagonal and Hermitian  we define
\begin{align}
    x_1 \equiv  X_{1}\,^1
    =
    1+\left(c_{HH1}^{(6)}-c_{HH2}^{(6)}\right)\,\frac{v^2}{\Lambda^2}\,, 
    \qquad
    x_2\equiv X_{2}\,^2
    =
    1+\left(c_{HH1}^{(6)}+c_{HH2}^{(6)}\right)\,\frac{v^2}{\Lambda^2} \,,
\end{align}
and at tree-level in the SM,  $x_1=x_2=1$.
For convenience  we  define the combinations, 
\begin{align}
    \alpha
    &=
    g_2 \left(1+c_{WW1}^{(6)}\,\frac{v^2}{\Lambda^2}+ c_{WW2}^{(8)}\,\frac{v^4}{\Lambda^4}\right)+g_1\, c_{WB}^{(6)}\, \frac{v^2}{\Lambda^2}\, ,\nn\\
    \beta
    &=
    g_1\,\left(1+c_{BB}^{(6)}\,\frac{v^2}{\Lambda^2}\right)+g_2\, c_{WB}^{(6)}\, 
    \frac{v^2}{\Lambda^2} \,,  \\ 
    \gamma 
    &=  g_2\, \left(1+c_{WW1}^{(6)}\,\frac{v^2}{\Lambda^2}\right) \, .\nn
\end{align}
Combining the above equations leads to 
\begin{align}
    \label{eq:evecsAgain}
    (\hat{g}^A_H)^{\,\,\,j=2}_{i=1} 
    = 
    \frac{x_1x_2}{2}\gamma\,(1,\, -i , \,0,\,0)\,,
    \qquad
    (\hat{g}^A_H)^{\,\,\,j=2}_{i=2} 
    =
    \frac{x_2^2}{2}(0,\, 0, \,-\alpha,\,\beta) \, ,
\end{align}
which are the building blocks of the gauge-field mass matrix of Eq.~\eqref{eq:mass-sq}. 
The vector boson masses are
\begin{align}
    \label{eq:wzmass}
    M_W^2 
    =
    \frac{x_1^2 x_2^2}{4}\gamma^2 v^2   
    \, , \qquad \qquad
    M_Z^2 
    =
    \frac{x_2^4}{4}(\alpha^2+\beta^2)\,v^2
    \,.
\end{align}
The mass states are given by, 
\begin{align}
    \tilde{V}_{\pm}
    &\equiv 
    \tilde V_1^{(*)}
    = \frac1{\sqrt2}\, (1,\,\mp i,\,0,\,0)\,,  
    \nn\\
    \tilde{V}_Z
    &\equiv -\tilde V_2
    = 
    (0,\,0,\,\cos\theta_w ,\,-\sin\theta_w)\, ,
    \\
    \tilde V_\gamma
    &\equiv
     (0,\,0,\,\sin\theta_w,\,\cos\theta_w) \,, \nonumber 
\end{align}
where 
\begin{align}
    \label{eq:thetaKX}
    \tan\theta_w= 
    \frac{\beta}{\alpha}
    \approx
    \frac{g_1}{g_2}
    \left[ 
    1
    +\left(c_{BB}^{(6)}- c_{WW1}^{(6)}
    +\left(\frac{g_2}{g_1}-\frac{g_1}{g_2}\right)\,c_{WB}^{(6)}
    \right)\frac{v^2}{\Lambda^2}
    -c_{WW2}^{(8)}\,\frac{v^4}{\Lambda^4}
    \right]
     \, ,
\end{align}
gives the weak mixing angle.
On the right-hand side of Eq.~\eqref{eq:thetaKX},  we work to leading order in the EFT expansion, but keep the $c^{(8)}_{WW2}$ term as it the leading contribution from the $W$-boson WFR. 
Based on the above analysis, we can identify the physical couplings in terms of the combinations
$\alpha$, $\beta$ and $\gamma$,
\begin{align}
\bar{g}_2 =\gamma\,,~~ 
 \bar{g}_Z\,\cos\theta_w = \alpha \,,~~ 
 \bar{g}_Z\,\sin\theta_w= \beta \,.
\label{eq:re_couplings}
\end{align}
Note that the $x_i$'s, which correspond to the scalar WFRs,  only appear in the masses, but not in the normalized eigenvectors $\tilde V_i$ which determine the mixing angle.
These scalar WFRs do appear in the $\rho$ parameter, 
\begin{align}
    \rho \equiv \frac{M_W^2}{M_Z^2\,\cos^2\theta_w}
    &=\frac{\gamma^2}{\alpha^2}\frac{x_1^2}{x_2^2}
    =\frac{\bar{g}_2^2}{\bar{g}_Z^2 \cos^2\theta_w}\frac{x_1^2}{x_2^2}
    \notag \\
    & \approx 1-4\,c_{HH2}^{(6)}\,\frac{v^2}{\Lambda^2} -2\,c_{WB}^{(6)}\frac{g_1}{g_2}\frac{v^2}{\Lambda^2} - 2\,c_{WW2}^{(8)}\,\frac{v^4}{\Lambda^4}  
    \, .
\end{align}
where in the last step we kept just the leading order pieces.
This expression clearly reflects the sources of custodial symmetry breaking in the SMEFT: 
(a)~the splitting between 
the charged and neutral Goldstone WFRs from $c_{HH2}^{(6)}$, 
(b)~the $c_{WW2}^{(8)}$ contribution to the 
$W$-bosons WFRs, 
and (c) the $W-B$ mixing $c_{WB}^{(6)}$. For $g_1=0$ the $B$ boson remains massless so $c_{WB}^{(6)}$ does not enter the $\rho$ parameter.

\section{Low-energy couplings}
\label{sec:LE_coup}
To derive the low-energy couplings we can rely either on the transverse or on the longitudinal amplitudes. 
We will use the latter. 
Consider first the $W^+$ couplings to SU(2)$_W$-doublet fermions.
The relevant massless $\bar{F}F'h G^+$ amplitude is,
\begin{align}
     \A \left( \bar{F}^{i}(p_1), F'_{j}(p_2) , h(p_3),G^+(p_4) \right)
     =
     \frac{1}{\sqrt{2}}\bigg[ 
     &-2\frac{\left(\hat{g}^A_{FV} \right)_i\,^j
    \left(\hat{g}^A_{H} \right)^{\,\,\,\ell=1}_{k=2}}{(p_1+p_2)^2}\,\delta_{FF'} \notag\\
    &\quad+
    \frac{1}{\Lambda^2}(\hat{c}_{FF'HH})_{i\ k=2}^{\ j \ \ell=1}
    \bigg]\, [132\rangle\,,\label{eq:FFhG}
\end{align}
where the second term is the SMEFT contact term of Eq.~\eqref{eq:ffHHgeneral}, and the
first term is the factorizable tree-level SM contribution with the SMEFT-corrected gauge couplings;
the $HHV$ coupling $\hat{g}^A_{H}$ and the fermion-fermion-vector coupling $\hat{g}^A_{FV}$.
For a fermion $\psi$, this coupling is given by,
\begin{align}
    \hat{g}^A_{\psi V} = K^{AB}\, g^B_{\psi V}\,,
\end{align}
with $g^b_{\psi V}=g_2\, \sigma^b/2$ and $g^4_{\psi V}= g_1\, Y_{\psi}$.
Matching the amplitude of Eq.~\eqref{eq:FFhG} to the longitudinal component of the low-energy $W$-amplitude as
described in Appendix~\ref{app:kin} one gets,
\begin{align}
    \frac{1}{M_W}(C^{W^+}_{FF'})_i\,^j
    = \frac{v}{\sqrt{2}}\bigg(-\frac{\left(\hat{g}^A_{FV} \right)_i\,^j
    \left(\hat{g}^A_{H} \right)^{\,\,\,\ell=1}_{k=2}}{M_W^2}\,\delta_{FF'}
    +
    \frac{1}{2\Lambda^2}(\hat{c}_{FF'HH})_{i\ k=2}^{\ j \ \ell=1}
    \bigg) \,.\label{eq:CWMW}
\end{align}
As we saw in Section~\ref{sec:mass-spur-XK},  $\left(\hat{g}^A_{H} \right)^{\,\,\,\ell=1}_{k=2}$ is nonzero only for $A=1,2$,
so only these terms survive in Eq.~\eqref{eq:CWMW}. 
Substituting the results for $K^{AB}$ and $\hat{g}^A_{H}$ from Sections~\ref{sec:SpuStrucuture} and~\ref{sec:mass-spur-XK} we find,
\begin{align}
    (C^{W^+}_{FF'})_i\,^j
    &
    = -\frac{\bar{g}_2}{\sqrt{2}}\left[
    \delta_{FF'}
    -
    \frac{v^2}{2\Lambda^2} \left(
    c^{(6)}_{FF'2}-ic^{(8)}_{FF'4}\,\frac{v^2}{\Lambda^2}
    \right)
    \right]\frac{(\sigma^+)_i\,^j}{2} \,,
\end{align}
where $\bar g_2$ is defined in Eq.~\eqref{eq:re_couplings}. 

The $W^+$ couplings to SU(2)$_W$ singlet quarks  are  given by Eq.~\eqref{eq:udssRH1-full}
with $k\neq \ell$, yielding
\begin{align}
    \frac1{M_W}\, C^{W^+}_{ud}
    =
    -\frac{v}{\sqrt{2}}\,(x_1x_2)^{-1}\, \frac{c^{(6)}_{ud}}{2\Lambda^2}\,,
\end{align}
and therefore,
\begin{align} \label{eq:RH-WcouplingNEW}
    C^{W^+}_{ud}
    =
    -\frac{\bar{g}_2}{\sqrt{2}}\, c^{(6)}_{ud}\,\frac{v^2}{4\Lambda^2}\,.
\end{align}
Note that there is no SM contribution to Eq.~\eqref{eq:RH-WcouplingNEW}, and that it has no leptonic counterpart since there are no right-handed neutrinos in the SMEFT. 

We can repeat the above matching for the $Z$ couplings. Since the universal contributions are more cumbersome
in this case we relegate the derivation to the Appendix. In Appendix~\ref{app:univ},  we perform this derivation in two
different ways: first using the transverse amplitudes and then using the longitudinal amplitudes.
The  $Z$ couplings to SU(2)$_W$-doublet fermions are obtained from  Eqs.~\eqref{eq:ffssLH1group} and \eqref{eq:Auniv} for $k=\ell=2$,
\begin{align}
    \frac{1}{M_Z}(C^{Z}_{FF'})_i\,^j
    &=
    v\,\left( \frac{\left(\hat{g}^A_{FV} \right)_i\,^{j}
    \left(\hat{g}^A_{H} \right)^{\,\,\,\ell=2}_{k=2}}{M_Z^2}\,\delta_{FF'}
    -
    \frac{1}{2\Lambda^2}(\hat{c}_{FF'HH})_{i\ k=2}^{\ j \ \ell=2}\right)\,,
\end{align}
giving,
\begin{align}
    (C^{Z}_{FF'})_i\,^j
    &
    =- \bar{g}_Z\,\left(\frac{1}{2}(\sigma_3)_i\,^j -\bar{s}^2\, Q_{F_i}\delta_i\,^j\right)\,\delta_{FF'}\notag
    \\
    &\quad+\bar{g}_Z\, \frac{v^2}{2\Lambda^2}\,\left(\left(c^{(6)}_{FF'2}+ c^{(8)}_{FF'3}\,\frac{v^2}{\Lambda^2}
    \right)\,\frac{1}{2}(\sigma^3)_i\,^j
    -
    c^{(6)}_{FF'1}\,\frac{1}{2}\delta_i\,^j\right)\,,
\end{align}
where $\bar g_Z$ and $\bar s^2$ are defined in Appendix \ref{app:univ}. 
Similarly, for SU(2)$_W$ singlet fermions, we obtain
\begin{align}
    \frac{1}{M_Z}C^{Z}_{ff'}
    &=
    v\, \left(\frac{\hat{g}^A_{fV} 
    \left(\hat{g}^A_{H} \right)^{\,\,\,\ell=2}_{k=2}}{M_Z^2}\,\delta_{ff'}-\frac{1}{2\Lambda^2}(\hat{c}_{ff'HH})_{k=2}^{\ \ell=2}\right)
    \,,
\end{align}
and therefore,
\begin{align}
    C^{Z}_{ff'}=\bar{g}_Z \bar{s}^2 Q_f\,\delta_{ff'}
    -
    \bar{g}_Z\,\frac{v^2}{4\Lambda^2}\, c^{(6)}_{ff'1}\,.
\end{align}
The photon couplings to fermions are derived from the transverse amplitudes and are given in Appendix~\ref{app:photon}.
These results are in agreement with those of Ref.~\cite{Helset:2020yio}.
\\
\\
We can now discuss the implications of these results.
Note first that the SU(2)$_W$ textures of the $W$ and $Z$ coupling are fully determined by the dimension-eight SMEFT, with no modifications beyond dimension-eight.
Furthermore, as expected, the number of observable left-handed couplings is larger than the number of theory inputs. 
For simplicity, we restrict our discussion now to a single generation,
and denote the left-handed quark couplings as $C^{W^+}_{UD}\equiv(C^{W^+}_{QQ})_1\,^2$, $C^Z_{UU}\equiv(C^Z_{QQ})_1\,^1$, $C^Z_{DD}\equiv(C^Z_{QQ})_2\,^2$, and the left-handed lepton couplings as $C^{W^+}_{\nu E}\equiv(C^{W^+}_{LL})_1\,^2$, $C^Z_{\nu\nu}\equiv(C^Z_{LL})_1\,^1$ and $C^Z_{EE}\equiv(C^Z_{LL})_2\,^2$. 

The $Z$-couplings are real. 
The measured quantity is the absolute value of each coupling, but since the leading contribution is the SM contribution, the sign of the coupling is also known. 
Thus, we can extract the dimensionless  three-point couplings with no sign ambiguity,
\begin{align} \label{eq:LHfZcouplings}
    C^Z_{UU/DD}
    =
    -\bar{g}_Z\left(\pm \frac{1}{2}-\bar{s}^2\, Q_{U/D}\right)
    -\frac{\bar{g}_Zv^2}{4\Lambda^2}\,\left(
    c^{(6)}_{QQ1}\mp c^{(6)}_{QQ2}\mp c^{(8)}_{QQ3}\,\frac{v^2}{\Lambda^2}
    \right)\,.
\end{align}

The $W$ couplings are complex, but their imaginary parts arise only at dimension-eight (again, for a single generation, the SM contribution is real) from $c^{(8)}_{QQ4}$. 
Therefore,
\begin{align}
    \left|C^{W^+}_{UD}\right|
    =
    \left|-\frac{\bar{g}_2}{\sqrt{2}}+\frac{1}{2\sqrt{2}}\frac{\bar{g}_2 v^2}{\Lambda^2}\,c^{(6)}_{QQ2}\right|+\cO\left( \frac{v^8}{\Lambda^8} \right) \,,
\end{align}
and to an excellent approximation we can extract,
\begin{align}
    \label{eq:subt-w}
    C^{W^+}_{UD} 
    \approx 
    -\frac{\bar{g}_2}{\sqrt{2}}+\frac{1}{2\sqrt{2}}\frac{\bar{g}_2 v^2}{\Lambda^2}\,c^{(6)}_{QQ2}\, .
\end{align}
\\
Combining Eqs.~\eqref{eq:LHfZcouplings} and~\eqref{eq:subt-w}, we can remove the 
universal terms, and isolate a combination of $W$ and $Z$ quark and lepton couplings which is not corrected by the dimension-six SMEFT,
\begin{align} \label{eq:SUM_equal}
    \left({C}^{Z}_{UU}-C^{Z}_{DD}\right)-\left(C^{Z}_{\nu\nu}-C^{Z}_{EE}\right)-\sqrt{2}\,
    \frac{\bar{g}_Z}{\bar{g}_2}\, \left( C^{W^+}_{UD} - C^{W^+ }_{\nu E} \right)
    =
    \left(c^{(8)}_{QQ3} -c^{(8)}_{LL3}\right)\, \frac{\bar{g}_Zv^4}{2\Lambda^4}\,.
\end{align}
Here, $\bar{g}_Z$ and $\bar{g}_2$ are the physical couplings.
We can also rewrite the left-hand side in terms of the physical masses,
\begin{align}\label{eq:rel-m}
    \left({C}^{Z}_{UU}-C^{Z}_{DD}\right)-\left(C^{Z}_{\nu\nu}-C^{Z}_{EE}\right)-\sqrt{2}\,
    \frac{M_Z}{M_W}\, \left( C^{W^+}_{UD} - C^{W^+ }_{\nu E} \right)
    ={\cal O}\left(\frac{v^4}{\Lambda^4}\right)\,.
\end{align}

We now turn to SM contributions to the spurion expansion.
Recall that so far we worked to leading order in the gauge coupling.
In particular, the low-energy couplings were determined via matching to the massless $\bar \psi \psi' H^\dagger H$ amplitude, where we keep the tree-level SM contribution with the SMEFT-corrected gauge coupling.
SM corrections on vector and scalar legs are effectively included in our analysis.
The tree-level effects of additional gauge coupling insertions on scalar and vector lines are taken into account, since these just generate the vector masses.
SM oblique corrections are captured by the matrices $X$ and $K$, with the spurion $\langle H\rangle/\Lambda$ replaced by $\langle H\rangle/v$.
Thus, the relations Eqs.~\eqref{eq:SUM_equal} and~\eqref{eq:rel-m} include these effects. These relations are not spoiled by QCD 1-loop contributions either, since these do not modify the spurion structue.

Pure SM effects can be treated in analogy with the SMEFT contributions.
For left-handed quarks, for instance, as we saw above, the tree-level SM contributes to the coefficients $c^{(6)}_{QQ1}$, $c^{(6)}_{QQ2}$. 
The remaining coefficients are generated by one-loop electroweak vertex corrections with additional Higgs insertions.

\section{Conclusions} 
\label{sec:conclusions}

Amplitude-based formulations of the SMEFT organize the SMEFT expansion as an expansion in amplitude contact terms with increasing numbers of external legs and kinematic structures.
Contact-terms with a given number of external-leg fields are spanned by a finite basis of kinematic structures, or Stripped Contact Terms~(SCTs), which  carry the external polarization information~\cite{Durieux:2020gip} (see also Ref.~\cite{Chang:2022crb}).
The full kinematic structure is then obtained by multiplying the SCT basis by Lorentz invariants. 
This kinematic expansion corresponds to the field- and derivative-expansion of Lagrangian formulations.

In this paper, we studied the Higgs field expansion of the physical low-energy amplitudes predicted by the SMEFT.
Each low-energy amplitude gets contributions from an infinite set of high-energy amplitudes with varying numbers of Higgses set to their VEVs. 
Just as in the case of SCTs, however, the $\Gew$ structure of the amplitude is spanned by a finite basis.
This is no surprise: both expansions are determined by the representations of the external legs under the relevant global symmetry. 
The number of SCTs is determined by the polarizations of the external legs, and the number of independent Higgs-spurion structures is determined by the $\Gew$ representations of the external legs.

The amplitude fomulation of the SMEFT allows for a simple spurion analysis in terms of the Higgs, underscoring the central role of the physical symmetries defining the theory: Lorentz on the one hand, and the internal SU(3)$_c\times \Gew$ on the other.
Here we used this to derive the $W$ and $Z$ masses and mixings, and the $\Gew$ structure of their three-point couplings to fermions.
The latter are proportional to a single SCT with no kinematic expansion.
It is simple to extend the spurion analysis to include additional Higgses.
Adding for example a triplet Higgs would result in an additional spurion $\lambda_T^a$ in Eq.~\eqref{eq:ffssLH1group}, with say, nonzero $a=2,3$ entries, which would modify the textures of the low-energy couplings and the physical spectrum.
Finally, since the spurion expansion captures the full $\Gew$, it may provide a way to distinguish the SMEFT from the HEFT.

\section*{Acknowledgments}
We thank Adam Martin and Gauthier Durieux for comments on the draft.
The work of Y.~Shadmi was performed in part at The Aspen Center for Physics, which is supported by National Science Foundation grant PHY-2210452, and at the KITP,  which is supported in part by grant NSF PHY-2309135 to the Kavli Institute for Theoretical Physics (KITP).
Research supported by ISF Grants No.~1002/23 and 597/24, by NSF-BSF Grant No.~2020-785, and by BSF Grants No.~2024-169 and No.~2024091.
Y.~Soreq thanks CERN-TH for the scientific associateship. 

\appendix

\section{Matching: kinematics}
\label{app:kin}
In this Appendix, we summarize the matching of high-energy and low-energy amplitudes. 
For additional details, we refer the reader to Ref.~\cite{Balkin:2021dko}. 
Consider the three-point amplitude involving two massless fermions $F$ and $F'$ (which we take to be left-handed for concreteness) and a massive vector $V$:
\begin{equation}\label{eq:mass3pt}
    \M(\bar{F}(p_1), F'(p_2), V(p_3)) = \frac{C}{M_V}\, [1{\bf3}]\anb{2{\bf3}}\,.
\end{equation}
where $C$ is a dimensionless coupling.  The bolded spinor structure implies symmetrization over the massive little-group indices, $[1{\bf3}]\anb{2{\bf3}}=[13^{\{I}]\anb{23^{J\}}}$, which gives $[13^{\{I}]\anb{23^{J\}}}$ $=[13^{I}]\anb{23^{I}}$ for $I=J$ and $[13^{\{I}]\anb{23^{J\}}}=\left([13^{I}]\anb{23^{J}}+[13^{J}]\anb{23^{I}}\right)/\sqrt{2}$ for $I\neq J$.
The spinors $3^I]$ and $3^I\rangle$, which encode the vector polarization, are defined by decomposing $p_3$ as $p_3=k+q$ with $k^2=q^2=0$. Note that $[qk]=\anb{kq}=M_V$.
For details see, e.g., Ref.~\cite{Durieux:2019eor}.
For a positive vector polarization, the massive amplitude of Eq.~\eqref{eq:mass3pt} can be written as,
\begin{equation}\label{eq:match+}
    \M(\bar{F}(p_1), F'(p_2), V_{+}(p_3)) =\frac{C}{M_V}\, [13^{1}]\anb{23^{1}}=C\,\frac{[1k]\anb{2q}}{\anb{kq}} = C\, \frac{[1k]^2}{[21]}\,,
\end{equation}
which is identical to the amplitude featuring a massless positive-helicity vector of momentum $k$,
\begin{equation}\label{eq:match+}
    \A(\bar{F}(p_1), F'(p_2), V_+(k)) = C\, \frac{[1k]^2}{[21]}\,.
\end{equation}
Throughout this appendix, the polarization or helicity of the vector $V$ is indicated by the lower subscript ($\pm$ and 0).
For longitudinal vector polarization, the amplitude in Eq.~\eqref{eq:mass3pt} can be written as,
\begin{equation}\label{eq:match0}
    \M(\bar{F}(p_1), F'(p_2), V_0(p_3)) =\frac{C}{M_V}\, [13^{\{1}]\anb{23^{2\}}}= \sqrt{2}C M_V\,\frac{[1k]\anb{2k}}{[kq]\anb{qk}}\,.
\end{equation}
This has the same form as the tree-level SM massless fermion-fermion-Higgs-Higgs amplitude,
\begin{equation}\label{eq:a0}
    \A(\bar{F}(p_1), F'(p_2), G(k), h(q)) = c_{UV}\, \frac{[1k]\anb{2k}}{[kq]\anb{qk}}\,,
\end{equation}
where $c_{UV}$ is dimensionless. 
Matching Eqs.~\eqref{eq:match0} and \eqref{eq:a0} requires a scale, since it relates a three-point amplitude and a four-point amplitude.
It is natural to associate this scale with the Higgs VEV,
\begin{equation}\label{eq:vevmat}
   \sqrt{2}C M_V = \Vev\, c_{UV}\,. 
\end{equation}
A priori, the right-hand side of Eq.~\eqref{eq:vevmat} can feature some numerical constant, but this can be absorbed in $\Vev$.
In fact, Eq.~\eqref{eq:vevmat} can serve as the definition of the Higgs VEV.
Explicit calculations confirm that the identification in Eq.~\eqref{eq:vevmat} is consistent for various amplitudes, at least to leading order in $\Vev$~\cite{Balkin:2021dko,Liu:2023jbq}.

In particular, for  $V=W^+$ for example, i.e., $G=G^+$, the tree-level SM  couplings are  $C=g_2$, $c_{UV}=-(g_2)^2/\sqrt{2}$, and $M_W=g_2v/2$.
For the SMEFT, the high-energy amplitude is instead of the form,
\begin{equation}\label{eq:asmeft}
    \A(\bar{F}(p_1), F'(p_2), G(k), h(q)) = -\frac{c^{(6)}_{UV}}{\Lambda^2}\, 
    [1k]\anb{2k}\,,
\end{equation}
and matching to Eq.~\eqref{eq:match0},
\begin{align}
    C= \frac{vM_V}{\sqrt2}\frac{c^{(6)}_{UV}}{\Lambda^2}\propto g \frac{v^2 c^{(6)}_{UV}}{\Lambda^2}\,.
\end{align}
Matching at higher orders in $\Vev$ may involve additional combinatorial factors, but these do not affect the spurion expansions, because such factors can be absorbed in the coefficients of the expansion.

It is useful to write the SM tree-level amplitude in terms of the coefficients of Eq.~\eqref{eq:ffssLH1group},
\begin{align}
    \label{eq:ffssLH1-sm}
     \A^{\text{SM}} \left( \bar{F}^{i}(p_1), F'_{j}(p_2) , H^{\dagger\,k}(p_3),H_{\ell}(p_4) \right)
    &= 
  2\,\Big[ c_{FF'1}^{(\text{SM})} \,\delta_{i}^{\,\,\,j} \delta_{k}^{\,\,\,\ell} \notag \\
  & \qquad \qquad
    +c_{FF'2}^{(\text{SM})}\,(\sigma^a)_{i}^{\,\,\,j} (\sigma^a)_{k}^{\,\,\,\ell}\Big]
    \frac{\sab{132}}{(p_1+p_2)^2} \,,
\end{align}
with $c_{FF'1}^{(\text{SM})}= -\delta_{FF'}\,Y_F\, g_1^2/2$, $c_{FF'2}^{(\text{SM})}= -\delta_{FF'}\,g_2^2/4$.
Then the $W$ coupling is obtained for $k\neq \ell$ with $(p_1+p_2)^2=M_W^2$, and the $Z$ coupling is obtained for $k = \ell=2$ with $(p_1+p_2)^2=M_Z^2$.

\section{Universal contributions to the low-energy couplings}
\label{app:univ}
In this Appendix, we derive the universal contributions to the $W$- and $Z$-fermion  couplings, 
as well as the photon couplings.
In Appendix~\ref{app:trans} we use the transverse amplitudes to derive the universal contributions to the $W$ and $Z$ couplings, and in Appendix~\ref{app:long}
we repeat this using the longitudinal amplitudes, demonstrating the consistency of the two derivations.

\subsection{Universal corrections to $W$ and $Z$ couplings}
\label{app:WZcoupUNIV}

\subsubsection{Transverse components}
\label{app:trans}
The universal corrections to fermion-$W$ and fermion-$Z$ couplings can be determined by matching the transverse high-energy and low-energy amplitudes.
The relevant high-energy amplitudes are the three-point gauge couplings, including the SMEFT vector WFR effects and the rotation to the mass basis.
For left-handed fermions, the three-point gauge coupling contribution takes the form:
\begin{align}\label{eq:vectors_WFR}
    \mathcal{A}\left(\bar{\psi}(p_1),\psi(p_2), V^{A}_+(p_3)\right)
    =
    -\sqrt{2}\,(\hat{g}_{\psi V}^A)\,\frac{[13]^2}{[12]}\,,
\end{align}
where the SMEFT vector WFR-corrected gauge couplings are defined by $\hat{g}^A_{\psi V}=K^{AB}\, g^B_{\psi V}$, with the SM tree-level couplings given by $g^b_{\psi V}=g_2\, \sigma^b/2$ and $g^4_{\psi V}= g_1\, Y_{\psi}$. 
For right-handed fermions, $\tfrac{[13]^2}{[12]}$ is replaced by $-\tfrac{[23]^2}{[12]}$.
The $W^+$ couplings to left-handed fermions are then,
\begin{align}
    \left(C^{W^+}_{FF} \right)_{i}\,^j \bigg|_{\rm{univ.}} =-\frac{1}{\sqrt{2}} \left( \hat{g}^1_{FV} + i\hat{g}^2_{FV} \right)^{\,\,\,j}_{i}
    =-\frac{\bar{g}_2}{\sqrt{2}} \frac{(\sigma^{+})^{\,\,\,j}_{i}}{2}\,,\label{eq:trans_Wpl_univ}
\end{align}
where $\bar{g}_2=g_2\left(1+c^{(6)}_{WW1}\tfrac{v^2}{\Lambda^2}\right)$ denotes the renormalized SU(2)$_W$ gauge coupling.
There are no universal couplings of the $W$ to right-handed fermions.

By transforming Eq.~\eqref{eq:vectors_WFR} to the mass basis of the vector gauge bosons, one obtains the universal corrections to the $Z$ couplings, with the WFR effects already included in $\hat{g}_{\psi V}$.
The $Z$ couplings to left-handed fermions are then,
\begin{align}\label{eq:Z_left_univ_trans}
    \left(C^{Z}_{FF} \right)_i\,^j \bigg|_{\rm{univ.}} 
    = -\frac{\alpha\, (\hat{g}^3_{FV})_i\,^j - \beta\,(\hat{g}^4_{FV})_i\,^j}{\sqrt{\alpha^2+\beta^2}}=-\bar{g}_Z\left(\frac{1}{2}(\sigma_3)_i\,^j-\bar{s}^2\, Q_{F_i}\,\delta_i\,^j\right)\,,
\end{align}
where $\bar{g}_Z$ and $\bar{s}^2$ are defined as
\begin{align}\label{eq:Zuniv_L}
     \bar{g}_Z=\sqrt{g_1^2 +g_2^2}\,(1+\delta_{Z1})^{1/2} ,\qquad \bar{s}^2= \frac{g_1^2}{g_1^2+g_2^2}\,\left(\frac{1+\delta_{Z2}}{1+\delta_{Z1}}\right)\,,
\end{align}
with the SMEFT corrections defined by
\begin{align}
    \delta_{Z1}&=\frac{1}{g_1^2+g_2^2}\,\bigg[
    \left(2g_1^2\,c_{BB}^{(6)}+ 4 g_1 g_2\,c_{WB}^{(6)}
    +
    2g_2^2\, c^{(6)}_{WW1}\right)\, \frac{v^2}{\Lambda^2}+\bigg(g_1^2\,(c_{BB}^{(6)})^2\notag
    \\
    &+
    \left(g_1^2+g_2^2\right)\,(c_{WB}^{(6)})^2+
    g_2^2\,\left(2c_{WW2}^{(8)}+(c_{WW1}^{(6)})^2\right)+
    2g_1 g_2\, (c^{(6)}_{BB}+c^{(6)}_{WW1})\,c^{(6)}_{WB}\bigg)\,\frac{v^4}{\Lambda^4}\notag
    \\
    &+2c_{WW2}^{(8)}\left(g_1g_2\, c_{WB}^{(6)}+g_2^2\, c_{WW1}^{(6)}\right)\,\frac{v^6}{\Lambda^6}
    +
    g_2^2\, (c_{WW2}^{(8)})^2\,\frac{v^8}{\Lambda^8}\bigg]\,,
    \\
    \delta_{Z2}&=\left(
    2 c^{(6)}_{BB}+\frac{2g_2}{g_1}\,c_{WB}^{(6)}
    \right)\,\frac{v^2}{\Lambda^2}
    +
    \left((c_{BB}^{(6)})^2+(c_{WB}^{(6)})^2+\frac{g_2}{g_1}\,c_{WB}^{(6)}\left(
    c_{BB}^{(6)}+c_{WW1}^{(6)}
    \right)\right)\,\frac{v^4}{\Lambda^4}\notag
    \\
    &+ \frac{g_2}{g_1}\, c_{WB}^{(6)} c_{WW2}^{(8)}\,\frac{v^6}{\Lambda^6}\,.
\end{align}
Then, the electric charge is given by $Q_\psi=\sigma^3/2 +Y_{\psi}$.
Similarly, the $Z$ couplings to right-handed fermions take the form:
\begin{align}
C^{Z}_{ff}&\bigg|_{\rm{univ.}}  =-\frac{\alpha\, \hat{g}^3_{fV} - \beta\, \hat{g}^4_{fV}}{\sqrt{\alpha^2+\beta^2}}=\bar{g}_Z \bar{s}^2 Q_{f}\,.\label{eq:Z_trans_RH}
\end{align}

\subsubsection{Longitudinal components}
\label{app:long}
Including SMEFT corrections to the gauge couplings, we obtain the following universal corrections:
\begin{align}
&\A\left( \bar{F}^{i}(p_1), F_{j}(p_2) , H^{\dagger\,k}(p_3),H_{\ell}(p_4) \right)=-2 (\hat{g}_{FV}^A)_i\,^j (\hat{g}_{H}^A)_k\,^\ell\, \frac{[132\rangle}{(p_1+p_2)^2} 
\,.\label{eq:Auniv}
\end{align}
The $W^+$ coupling to SU(2)$_W$-doublet fermions is given by Eq.~\eqref{eq:Auniv} for $k=2$ and $\ell=1$,
\begin{align}\label{eq:Wuniv_L}
    \left(C^{W^+}_{FF} \right)_{i}\,^j \bigg|_{\rm{univ.}} 
    =-\frac{\bar{g}_2}{\sqrt{2}} \frac{(\sigma^{+})^{\,\,\,j}_{i}}{2}\,,
\end{align}
where the $W$ boson mass in Eq.~\eqref{eq:wzmass} is used.
This result agrees with Eq.~\eqref{eq:trans_Wpl_univ}.
Moreover, the $Z$ coupling is given by Eq.~\eqref{eq:Auniv} for $k=2$ and $\ell=2$,
\begin{align}
    \left(C^{Z}_{FF} \right)_i\,^j \bigg|_{\rm{univ.}} 
    = -\bar{g}_Z\left(\frac{1}{2}(\sigma_3)_i\,^j-\bar{s}^2\, Q_{F_i}\,\delta_i\,^j\right)\,,
\end{align}
where $\bar{g}_Z$ and $\bar{s}^2$ are defined in Eq.~\eqref{eq:Zuniv_L}.
This result is also consistent with Eq.~\eqref{eq:Z_left_univ_trans}.

Similarly, the universal corrections to the couplings to right-handed fermions are given by
\begin{align}\label{eq:Auniv_f}
&\A\left( \bar{f}(p_1), f(p_2) , H^{\dagger\,k}(p_3),H_{\ell}(p_4) \right)=-2 (\hat{g}_{fV}^A) (\hat{g}_{H}^A)_k\,^\ell\, \frac{\langle 132]}{(p_1+p_2)^2} 
\,.
\end{align}
The $Z$ coupling is derived from Eq.~\eqref{eq:Auniv_f} for $k=2$ and $\ell=2$,
\begin{align}
    C^{Z}_{ff}&\bigg|_{\rm{univ.}}  =\bar{g}_Z \bar{s}^2 Q_{f}\,,
\end{align}
which is consistent with Eq.~\eqref{eq:Z_trans_RH}.

\subsection{Photon couplings}
\label{app:photon}
By transforming Eq.~\eqref{eq:vectors_WFR} to the mass basis of the vector gauge bosons, one obtains the photon couplings in Eq.~\eqref{eq:photon_coup}:
\begin{align}
    e Q_{F_i}\,\delta_i\,^j &= \frac{\beta\, (\hat{g}^3_{FV})_i\,^j + \alpha\,(\hat{g}^4_{FV})_i\,^j}{\sqrt{\alpha^2+\beta^2}}\,,\quad e Q_{f} =\frac{\beta\, \hat{g}^3_{fV} + \alpha\, \hat{g}^4_{fV}}{\sqrt{\alpha^2+\beta^2}}\,.
\end{align}
By substituting the explicit forms of $\alpha$, $\beta$, and the SMEFT vector WFR effects, we find that the electric charge $e$ is universal and can be expressed 
at leading order in the EFT expansion as
\begin{align}\label{eq:e_approx}
    e &\approx \frac{g_1g_2}{\sqrt{g_1^2+g_2^2}}
    \Bigg(
    1 + \left(\frac{g_1^2}{g_1^2+g_2^2}\, \left(c_{WW1}^{(6)}+c_{WW2}^{(8)}\,\frac{v^2}{\Lambda^2}\right)
    +
    \frac{g_2^2}{g_1^2+g_2^2}\, c_{BB}^{(6)}
    -
    \frac{2g_1 g_2}{g_1^2+g_2^2}\, c^{(6)}_{WB}\right)\, \frac{v^2}{\Lambda^2}
    \Bigg)\,.
\end{align}

\section{General spurion expansion of $HHV$}
\label{app:mass-spur-gen}
%
In this Section, we provide the general spurion expansion of $HHV$ of Eq.~\eqref{eq:Hgeneral}.
We first mention some properties of the spurion coefficients of Eq.~\eqref{eq:Hgeneral}.
Consider the complex-conjugate of the $HHV$ amplitude.
This corresponds to the amplitude with $1\leftrightarrow2$, $i\leftrightarrow j$ and flipped-helicity vector.
Using the relation between scalar-scalar-vector amplitudes for negative- and positive-helicity vectors (see, e.g., Ref.~\cite{Durieux:2019eor}),
\begin{equation}
    \A(S(p_1), S^\prime(p_2), V_+(p_3))=C \frac{[13][23]}{[12]}\,,~~~
    \A(S(p_1), S^\prime(p_2), V_-(p_3))=C \frac{\anb{13}\anb{23}}{\anb{12}}\,,
\end{equation}
it is easy to verify that $C$ is real-valued (for $S=H^{\dagger\,i}$ and $S'=H_j$ the coupling $(\hat{g}^A_H)_{i}\,^j$ is Hermitian).
Thus, the spurion coefficients of Eq.~\eqref{eq:Hgeneral} are real.

We note that $c_{HW1}$ is proportional to $g_2$, and the remaining couplings have pieces proportional to $g_2$
and pieces proportional to $g_1$.
For convenience, we express  $c_{HW1}$, $c_{HW2}$, $c_{HW3}$, $c_{HB1}$, $c_{HB2}$, in terms of  $c_{WW1}^{(6)}$, $c^{(8)}_{WW2}$, $c^{(6)}_{WB}$, $c_{BB}^{(6)}$, $c^{(6)}_{HH1}$ and $c^{(6)}_{HH2}$:
\begin{align}\label{eq:gen-xk-reln}
    c_{HW1}&=\frac{g_2}{2}\, \left(1+c_{WW1}^{(6)}\,\frac{v^2}{\Lambda^2}\right)
    \left(
    \left(1+c_{HH1}^{(6)}\,\frac{v^2}{\Lambda^2}\right)^2
    -
    \left(c_{HH2}^{(6)}\right)^2
    \left(\lambda^a_H\right)^2
    \right)\,,
    \\
    c_{HW2}&=\frac{g_1}{2}\, c_{WB}^{(6)}\,\left(
    \left(1+c_{HH1}^{(6)}\,\frac{v^2}{\Lambda^2}\right)^2
    +
    \left(c_{HH2}^{(6)}\right)^2
    \left(\lambda^a_H\right)^2
    \right)\notag
    \\
    &\quad+g_2\, \left(
    \left(1+c_{WW1}^{(6)}\,\frac{v^2}{\Lambda^2}\right)
    +
    c_{WW2}^{(8)} \left(\lambda^a_H\right)^2
    \right)\left(1+c_{HH1}^{(6)}\,\frac{v^2}{\Lambda^2}\right)
    c_{HH2}^{(6)}\,,
    \\
    c_{HW3}&=g_1\, c_{WB}^{(6)}\, \left(1+c_{HH1}^{(6)}\,\frac{v^2}{\Lambda^2}\right) c_{HH2}^{(6)}\notag
    \\
    &\quad+
    \frac{g_2}{2}\, \Bigg(
    2 \left(1+c_{WW1}^{(6)}\,\frac{v^2}{\Lambda^2}+c_{WW2}^{(8)}\, \left(\lambda^a_H\right)^2\right)\, \left(c_{HH2}^{(6)}\right)^2\notag
    \\
    &
    \qquad\qquad+
    c_{WW2}^{(8)}\, \left(
    \left(1+c_{HH1}^{(6)}\,\frac{v^2}{\Lambda^2}\right)^2
    -
    \left(c_{HH2}^{(6)}\right)^2
    \left(\lambda^a_H\right)^2
    \right)
     \Bigg)\,,
    \\
    c_{HB1}&=\frac{g_1}{2}\, \left(1+c_{BB}^{(6)}\,\frac{v^2}{\Lambda^2}\right)
    \left(
    \left(1+c_{HH1}^{(6)}\,\frac{v^2}{\Lambda^2}\right)^2
    +
    \left(c_{HH2}^{(6)}\right)^2
    \left(\lambda^a_H\right)^2
    \right)\notag
    \\
    &\quad+
    g_2\, c_{WB}^{(6)} \left(1+c_{HH1}^{(6)}\,\frac{v^2}{\Lambda^2}\right)c_{HH2}^{(6)}\left(\lambda^a_H\right)^2\,,
    \\
    c_{HB2}&=g_1\, \left(1+c_{BB}^{(6)}\,\frac{v^2}{\Lambda^2}\right)\left(1+c_{HH1}^{(6)}\,\frac{v^2}{\Lambda^2}\right) c_{HH2}^{(6)}\notag
    \\
    &\quad+
    \frac{g_2}{2}\, c_{WB}^{(6)}\, \left(
    \left(1+c_{HH1}^{(6)}\,\frac{v^2}{\Lambda^2}\right)^2
    +
    \left(c_{HH2}^{(6)}\right)^2\left(\lambda^a_H\right)^2
    \right)\,.
\end{align}
%

\bibliographystyle{JHEP}
\bibliography{spurion.bib}

\end{document}